\documentclass[floatfix,showpacs,amsmath,amssymb,letterpaper,groupaddresses,superscriptaddress]{article}
\usepackage{bm}
\usepackage{amsmath}
\usepackage{amssymb}
\usepackage{amsthm}

\usepackage{qcircuit}
\usepackage{latexsym}
\usepackage{epsfig}
\usepackage{hyperref}
\usepackage{graphicx}
\graphicspath{ {./images/} }
\usepackage{amsmath}
\usepackage{amssymb}
\usepackage{graphicx}
\usepackage{times}
\usepackage[section]{placeins}
\usepackage{subfigure}
\usepackage{caption}
\usepackage[dvipsnames]{xcolor}
\usepackage{lmodern}
\usepackage{float}

\def\be{\begin{equation}}
\def\ee{\end{equation}}
\def\ba{\begin{eqnarray}}
\def\ea{\end{eqnarray}}
\def\la{\langle}
\def\ra{\rangle}

\def\h{\hskip 1cm}

\def\lo{\longrightarrow}
\def\a{\alpha}

\begin{document}

	\vspace{4cm}
	\begin{center}{\Large \bf Protecting  unknown qubit states from decoherence of qubit channels by weak measurement}
		\vspace{2cm}
		
		Ozra Heibati$\footnote[1]{email: o.heibati@azaruniv.ac.ir}$, \h Azam Mani$\footnote[2]{email: mani.azam@ut.ac.ir}$,\h Esfandyar Faizi$\footnote[1]{email: efaizi@azaruniv.ac.ir}$,\h Vahid Karimipour$\footnote[3]{Corresponding Author, email: vahid.karimipour@gmail.com}$\h \\
		\vspace{1cm}
$^\text{1}$Department of Fundamental  Sciences, Azarbaijan Shahid Madani University, Tabriz, Iran.\\
     \mbox{$^\text{2}$Department of Engineering Science, College of Engineering,  University of Tehran, Tehran, Iran.}\\
      $^\text{3}$Department of Physics, Sharif University of Technology,  Tehran, Iran.\\

	\end{center}
	\hfill
	
	\begin{abstract}
		The problem of combating de-coherence by weak measurements has already been studied for the amplitude damping channel and for specific input states. We generalize this to a large  four-parameter family of qubit channels and for the average fidelity over all pure states.  As a by-product we classify all the qubit channels which have one  invariant pure state and show that the parameter  manifold of these channels is isomorphic to $S^2\times S^1\times S^1$ and contains many interesting subclasses of channels. The figure of merit that we use is the average input-output fidelity which we show can be increased up to $30$ percents in some cases, by tuning of the weak measurement parameter.
				\\
				
	\end{abstract}

\section{Introduction}
The most general quantum operation on a quantum state is given by a Completely Positive Trace-preserving map or CPT for short \cite{kruas}. Such a  map or quantum channel is defined by its Kraus representation as

\be
\rho\lo {\cal E}(\rho)=\sum_i A_i \rho A_i^\dagger,
\ee
where the Kraus operators $A_i$ can be interpreted as probabilistic  operators which change the input state $\rho$ to the state $\rho_i=\frac{1}{Tr(A_i\rho A_i^\dagger)} A_i\rho A_i^\dagger$ with probability $p_i=Tr(A_i\rho A_i^\dagger)$. Mathematically these are the only operations which preserve all the properties of a quantum state and include non-unitary dynamics and measurements. Among the quantum channels, the most important ones are qubit channels, i.e. those which act on qubit states. The reason for this preference is clear enough since qubit states are the most common and the most studied candidates for storing and manipulating quantum information \cite{kruas}. From a theoretical point of view, their importance derives from their simplicity of structure and their complete classification \cite{Mbeth1}. \\

\noindent A qubit state passing through a channel necessarily loses its fidelity with the input state and to make quantum information processing practical, various kinds of error correction methods have been developed to retrieve the original state by measuring the error syndrome and then making necessary corrections \cite{error1, error2}. In other schemes, it has been tried to find de-coherence free subspaces of channels and encode information in these subspaces \cite{lidar}. In \cite{quasi inversion, quasiD} one tries to find a channel which when appended to the original channel, increases the average input-output fidelity.\\

Here we are more concerned with a rather mathematical problem, which is more in line with the recent works related to weak measurements \cite{Aharonov}. These works generally try to increase the efficiency of a quantum information processing task, by augmenting the corresponding quantum channel by a pre- and a post-weak measurement. The role of the post weak measurement is to reverse \cite{revers1,revers2,revers3}, as much as possible without compromising the efficiency of the new protocol, the effect of the pre-weak measurement. Almost all of these protocols consider one particular channel or quantum information processing task and specific class of states. Examples include the effect of weak measurement on feedback control systems \cite{feedback1,feedback2}, on spin squeezing \cite{spin}, and on protection of coherence of qutrit states \cite{coherence1} against decoherence \cite{coherence2, state1, state2, state3, state4,GHZ,}. Also this strategy has been used to protect various types of quantum correlations in two and multi-particle systems \cite{ entanglesharing,correlations, correlation,storing, entangle1,entangle2,entangle3,entangle4,entangle5,entangle6, discord1,discord2,discord3} and also on 
quantum teleportation, quantum dense coding and quantum telecloning in the presence of noise \cite{teleportation, densecoding, telecloning}.\\

\noindent Most of the weak measurement protection methods produced so far, have been restricted to preserve  certain initial states against the amplitude damping channel and in rare cases,
some other typical channels. It is natural to ask if it is possible to use the weak measurements for overcoming against other noise sources and if it shows improvement for various states of a quantum system. 
We stress that our work here is restricted to qubit channels, there are many  other works on how to amplify signals in quantum phase  communication channels which are affected by phase diffusion \cite{paris1, paris2, paris3, paris4, paris5}.\\

\noindent In this paper, we look at this problem from a more general point of view and show how the de-coherence of a large class of qubit channels can be reduced by only a weak measurement.  Our generalization is two-fold. First, we do not consider a specific channel like the Amplitude Damping channel, but introduce a multi-parameter family of channels, which contains the amplitude damping channel as a special case. Second, we  do not focus on special cases of input states, but rather we deal with protecting all states on the average. This is measured by comparing the  average input-output fidelity before and after processing by the weak-measurement. Our basic idea is to consider all those channels which have one invariant pure state and base our weak measurement on this invariant state. It should be noted that in our strategy the recovering process is not needed and we have in fact found, but not reported in our paper, that  a post-measurement always reduces the average input-output fidelity.  
We show that by suitably tuning of the weak measurement strength, it is possible to increase the average input-output fidelity which depending on the channel can be around $30$ percents. Besides this a by-product of our work is a classification of all qubit channels which have one invariant pure states, whose parameter space is shown in figure (\ref{S2fig}).\\

\noindent The structure of the paper is as follows. First in section (\ref{channelclass}), we classify all  qubit channels which have one pure invariant state. This is a multi-parameter family of qubit channels which contains many channels, including the amplitude damping channel as a special case. This by itself is a byproduct of this research which has not been reported before.  Then in section (\ref{weak}), we introduce the processing by weak-measurement and the rationale behind this kind of processing followed by subsection (\ref{fidelity}) in which we calculate the average fidelity of channels both before and after processing and give a brief discussion of the efficiency of our protocol. Finally, we summarize our conclusions in section (\ref{conclusion}).\\

\section{Qubit channels with one invariant pure state}\label{channelclass}
In this section, we classify qubit channels which have one invariant pure state.
Note that any channel with one invariant pure state can be connected in a simple way to a qubit channel whose invariant state is the state $|0\ra$. Let ${\cal E}^{\lambda}$ be an arbitrary qubit channel with a pure invariant state $|\lambda\ra\la \lambda|$, then it is easy to show that the channel
\be\label{ES}
{\cal E}(\rho)=S{\cal E}^\lambda (S^{-1}\rho S)S^{-1},
\ee
has the invariant state $|0\ra=S|\lambda\ra$.  As we will see in the next section, once we learn how to improve the average fidelity of ${\cal E}$, it is straightforward to improve the average fidelity of ${\cal E}^\lambda$.
For this reason hereafter we focus our attention to channels of the mentioned type, i.e. to channels whose invariant state is $|0\ra\la 0|$.
Our aim is to  classify all channels with the invariant state $|0\ra\la 0|$. We do this in the following theorem:\\

\noindent {\bf Theorem:} Every qubit channel with one invariant pure state is unitarily, in the sence of (\ref{ES}), equivalent to a channel of the following form:

\be
{\cal E}(\rho)=\sum_{i=0}^2 A_i\rho A_i^\dagger,
\ee
with
\be\label{newK}
A_0=\left(\begin{array}{cc} 1&0\\ 0 &y_0\end{array}\right),\h A_{1}=	\left(\begin{array}{cc}0&x\sin\theta\\ 0 &0\end{array}\right), \ \  A_{2}=	\left(\begin{array}{cc}0&-x\cos\theta\\ 0 &-ye^{\frac{i\phi}{2}}\end{array}\right),\ \
\ee
where $y_0, y$ and $x$ are real parameters subject to the trace-preserving condition
\begin{equation}\label{completeness}
y_0^2+x^2+y^2=1.
\end{equation}

\noindent This is a four-parameter family of qubit channels which in view of our mathematical reduction exhausts all channels which have the state $|0\ra$ as their invariant state. The form of Kraus operators cannot be simplified or their numbers cannot be reduced further.\\
\noindent Under the action of such channels, any state $\rho=\left(\begin{array}{cc}\rho_{00} & \rho_{01} \\ \rho_{10} & \rho_{11}\end{array}\right)$ transforms to
\be\label{general channel}
{\cal E}(\rho)=\left(\begin{array}{cc}\rho_{00}+x^2\rho_{11} & y_0\rho_{01}+ xye^{-i\frac{\phi}{2}}\cos\theta\rho_{11}\\ y_0\rho_{10}+xye^{i\frac{\phi}{2}}\cos\theta\rho_{11} &  (1-x^2)\rho_{11}\end{array}\right).
\ee
Before presenting the proof, let us elaborate on the properties of this channel and a few special subclasses.\\
\noindent The relation (\ref{completeness}) represents the equation of a 2-sphere $S^{2}$ and then the manifold of all such channels is given by ${\cal A}=S^{2}\times S^{1}\times S^{1}$. The average fidelity however, as we will see later, depends only on the parameters of $S^{2}$.\\

\begin{figure}
	\centering
	\includegraphics*[scale=0.4]{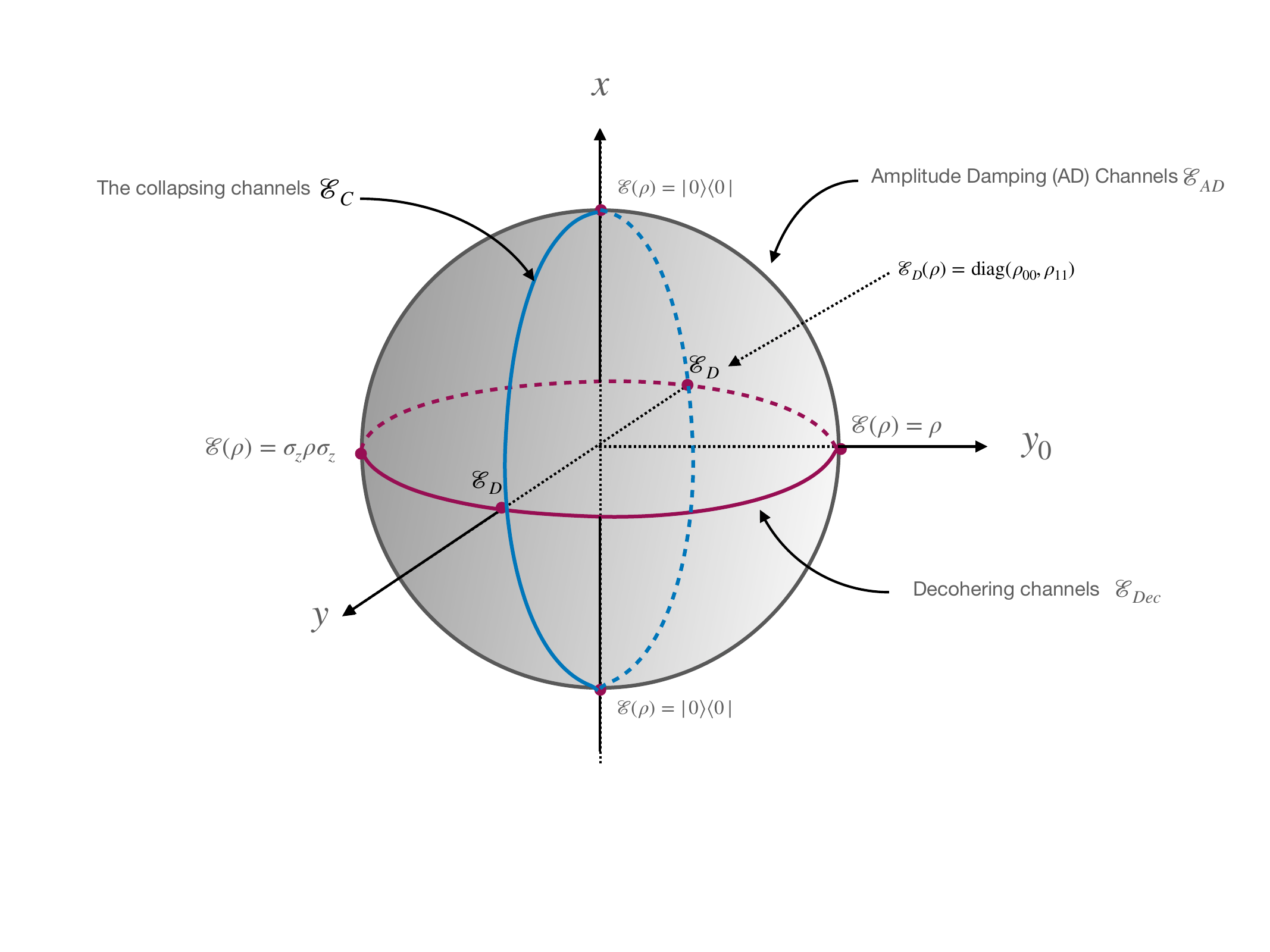}
	\vspace*{-3cm}
	\caption{The parameter space of the channels defined in (\ref{newK}). The points $(y_0,x,y)$ lie on a sphere $S^2$ and in each point two other continuous parameters $\theta$ and $\phi$ define a family of channels. Each of the two meridians and the equator correspond to well-known families of channels with two Kraus operators.  There are six points on the sphere which  correspond to  specific channels with no parameters.  }\label{S2fig}
	\vspace*{0cm}
\end{figure}

\noindent It turns out that this family of channels has a very rich structure and the manifold of channels contains many interesting families of channels as we will now see. Special members of these channels are those which have two orthogonal invariant stats, i.e. both $|0\ra\la 0|$ and $|1\ra\la 1|$ are invariant. This extra demand leads to the requirement $x=0$. Under this condition, the  Kraus operator $A_{1}$ vanishes and we are left with two Kraus operators pertaining to dephasing.  One can also show that there is no qubit channel with two non-orthogonal invariant states \cite{Mbeth1}.
Figure (\ref{S2fig}) indicates the parameter space of the family of qubit channels which have at least one invariant pure state. Each channel of this family is parameterized by a point with coordinates $(y_0,x,y)$ which  lies on a sphere $S^2$ and two parameters $\theta$ and $\phi$. It is interesting to note several special cases of such qubit channels:\\

$\bullet$ On the meridian $y = 0$, where  $y_0=\sqrt{1-x^2}$, the two Kraus operators $A_1$ and $A_2$ will be proportional and their effect can be combined to one single Kraus operator pertaining to the Amplitude Damping (AD) channel, namely
${\cal E}_{AD}(\rho)=\left(\begin{array}{cc}\rho_{00}+x^2\rho_{11}& y_0\rho_{01} \\ y_0\rho_{10} & (1-x^2)\rho_{11}\end{array}\right).$ Here $x$ plays the role of damping parameter. \\

$\bullet$ On the equator $x = 0$, the Kraus operator $A_1$ vanishes and the channel becomes the de-cohereing  channel, ${\cal E}_{\rm Dec}(\rho)=\left(\begin{array}{cc}\rho_{00}&y_0\rho_{01} \\ y_0\rho_{10}& \rho_{11}\end{array}\right)$, the amount of de-cohering depends on the magnitude of $y_0$, \\

$\bullet$ On the meridian $y_0=0$, where  $y=\sqrt{1-x^2}$, the output state (\ref{general channel}) can be written as
${\cal E}_C(\rho)=\rho_{00}|0\ra\la 0|+\rho_{11}(1-x^2)\sin^2\theta|1\ra\la 1|+\rho_{11}(\cos^2\theta+x^2\sin^2\theta)|\phi\ra\la\phi|$, where $|\phi\ra=\frac{1}{\sqrt{\cos^2\theta+x^2\sin^2\theta}}(x|0\ra+\sqrt{1-x^2}e^{i\frac{\phi}{2}}\cos\theta|1\ra)$.  Therefore this channel collapses the state into a mixture of three fixed states independent of the input: $|0\ra$, $|1\ra$ and the state $|\phi\ra$.\\

$\bullet$ Finally as shown in figure (\ref{S2fig}), special points on the sphere correspond to the identity channel, the unitary channel $\rho\lo\sigma_z\rho\sigma_z$, the collapsing channel $\rho\lo |0\ra\la 0|$ and completely dephasing channel $\rho\lo {\cal E}_D(\rho)=\left(\begin{array}{cc}\rho_{00}& 0 \\ 0 & \rho_{11}\end{array}\right)$ . \\

\noindent We are now ready to present the proof of the theorem. \\

\noindent {\bf Proof:}
 We first note that by suitable unitary transformations, the number of Kraus operators of a qubit channel can be reduced to at most four. In this minimal representation, a qubit  channel can be written as
\be
{\cal E}(\rho)=\sum_{i=0}^3 K_i \rho K_i^\dagger.
\ee
In view of (\ref{ES}), we take the invariant state to be $|0\ra\la 0|$ and $K_i=\left(\begin{array}{cc} k_i&l_i\\ m_i &n_i\end{array}\right)$. From ${\cal E}(|0\ra\la 0|)=|0\ra\la 0|$, we have
\be\label{kmat}
\sum_i \left(\begin{array}{cc} |k_i|^2&k_im_i^*\\ k_i^*m_i &|m_i|^2\end{array}\right)=\left(\begin{array}{cc} 1&0\\ 0 &0\end{array}\right),
\ee
from which we obtain that $m_i=0\ \ \forall\ i$ and hence the Kraus operators are of the form
\be\label{KK}
K_i=\left(\begin{array}{cc} k_i&l_i\\ 0 &n_i\end{array}\right).
\ee
From (\ref{kmat}), we find
\be
\sum_{i=0}^{3} |k_i|^2=1.
\ee
Now we use the freedom of choosing Kraus operators by making linear unitary combinations of them and changing $K_i$'s to $A_i$'s according to
\be
A_i=\sum_j U_{i,j}K_j
\ee
where $U$ is a unitary matrix. Considering the above equation element-wise, we find that due to (\ref{KK})
\be
(A_i)_{1,0}=\sum_j U_{i,j} (K_j)_{1,0}=0,
\ee
and
\be
a_i:=(A_i)_{0,0}=\sum_j U_{i,j} (K_j)_{0,0}=\sum_j U_{i,j}k_j.
\ee
This means that with a suitable unitary matrix $U$, we can rotate the vector ${\bf k}=(k_0, k_1, k_2, k_3)$ to ${\bf a}=(1,0,0,0).$ The new Kraus operators are of the form
\be
A_0=\left(\begin{array}{cc} 1&x_{0}\\ 0&y_0\end{array}\right)\ \ , \ \ A_{i\ne 0}=\left(\begin{array}{cc} 0&x_i\\ 0&y_i\end{array}\right).
\ee

\noindent Imposing the trace-preserving property on these new Kraus operators, we find the following conditions
\be\label{completennesschannel}
x_{0}=0,\qquad|y_0|^2+\overline{{\bf x}}\cdot {\bf x}+\overline{{\bf y}}\cdot {\bf y}=1,
\ee
where ${\bf x}=(x_1,x_2,x_3)$ and ${\bf y}=(y_1,y_2,y_3)$ are complex vectors.
Furthermore, we note that
if ${\cal E}$ has the invariant state $|0\ra\la 0|$, then the channel $\tilde{{\cal E}}(\rho)=T{\cal E}(\rho) T^\dagger$, where $T=\left(\begin{array}{cc}1&0\\0&e^{i\phi}\end{array}\right)$, has also the same invariant state. This will allow us to take the parameter $y_0$ to be real.
To make further simplification, let
\be
|{\bf x}|=x,\h |{\bf y}|=y,\h \overline{{\bf x}}\cdot {\bf y}=xye^{i\phi}\cos\theta
\ee
One can now use a further unitary transformation in the form
\be
U'=\left(\begin{array}{cc}1&0\\ 0 & \Omega\end{array}\right),
\ee
where $\Omega\in SU(3)$ and transforms the two complex vectors ${\bf x}$ and ${\bf y}$ into the following form
\be
{\bf x}=xe^{-\frac{i\phi}{2}}(\cos\frac{\theta}{2},\  \sin\frac{\theta}{2},\ 0)\ \ \ ,\ \ \ {\bf y}=ye^{\frac{i\phi}{2}}(\cos\frac{\theta}{2},\  -\sin\frac{\theta}{2},\ 0).
\ee
This means that the  form of the Kraus operators (which are now 3 instead of 4) is now given by
\be
A_0=\left(\begin{array}{cc} 1&0\\ 0 &y_0\end{array}\right),\h A_{1}=	\left(\begin{array}{cc}0&xe^{-\frac{i\phi}{2}}\cos\frac{\theta}{2}\\ 0 &ye^{\frac{i\phi}{2}}\cos\frac{\theta}{2}\end{array}\right), \ \  A_{2}=	\left(\begin{array}{cc}0&xe^{-\frac{i\phi}{2}}\sin\frac{\theta}{2}\\ 0 &-ye^{\frac{i\phi}{2}}\sin\frac{\theta}{2}\end{array}\right).
\ee
Finally using the freedom of a rotation between the Kraus operators
$$A_0\lo A_0,\h A_1\lo \sin\frac{\theta}{2}A_1+\cos\frac{\theta}{2}A_2,\h  A_2\lo -\cos\frac{\theta}{2}A_1+\sin\frac{\theta}{2}A_2,$$
followed by a unitary transformation ${\cal E}(\rho)\lo \omega {\cal E}(\omega^\dagger \rho \omega)\omega^\dagger$ with
$\omega=\left(\begin{array}{cc} e^{i\phi/2}&0\\ 0 &1\end{array}\right)$ which is equivalent to $A_i\lo \omega A_i \omega^\dagger$,  leads to the form of Kraus operators given in (\ref{newK}).
This completes the proof. \qquad $\blacksquare$\\

\section{Protecting a qubit state by weak-measurement}\label{weak}

\begin{figure}
\centering
\includegraphics*[scale=0.4]{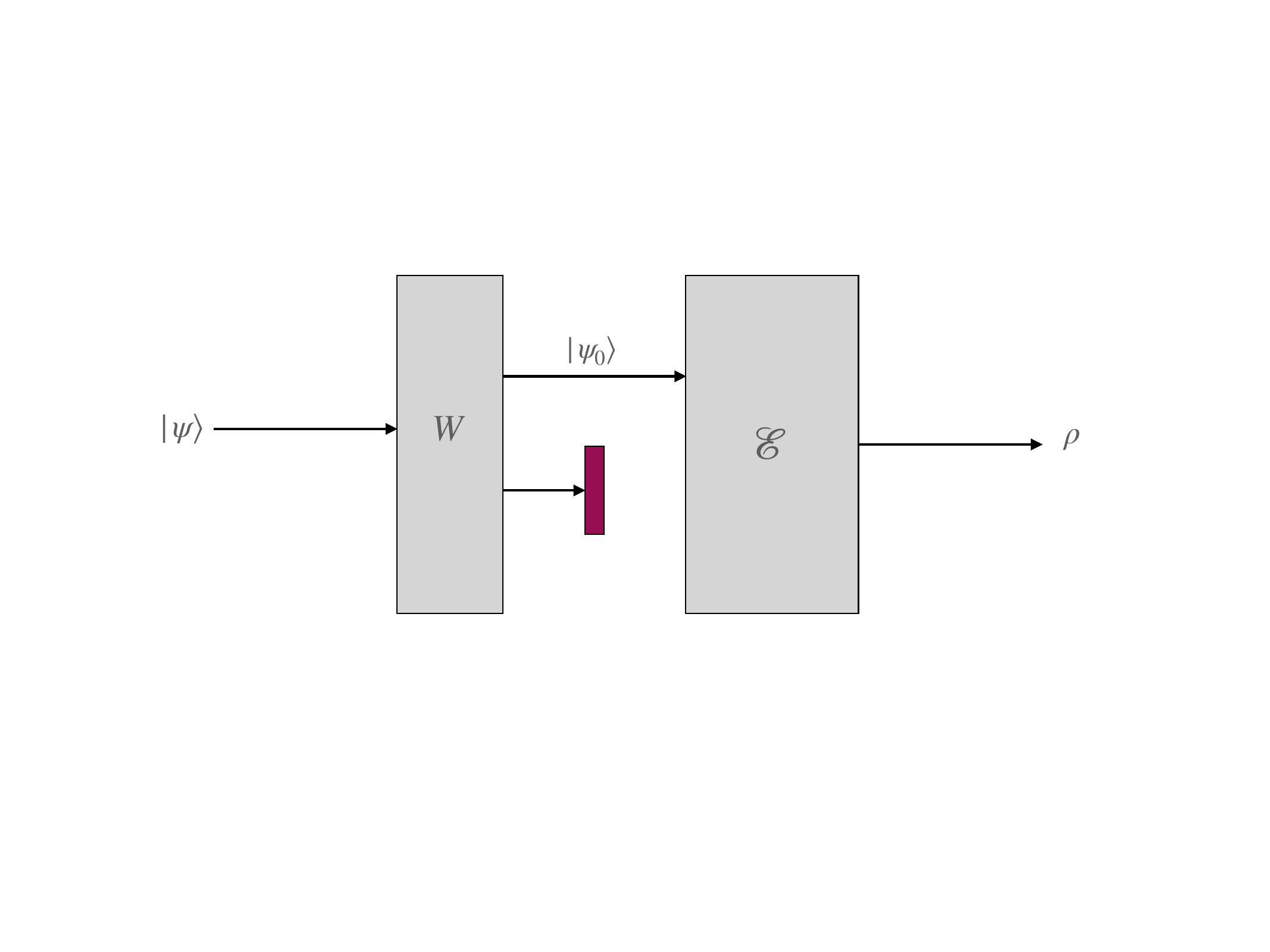}
\vspace*{-4cm}
\caption{Schematic of the state protection protocol. In order to decrease the decoherence due to an arbitrary qubit channel, $\cal E$, -(increase its average input-output fidelity) we present a protocol by using pre-processing. $W_{0}$ and $W_{1}$ are the pre-measurement operations. $W_0$ slightly bends the input state toward the invariant state and the output of $W_1$ which is orthogonal to the invariant state is discarded. The parameter of the  partial collapse measurement is optimally chosen to attain the best result.  }\label{setup}
\vspace*{0cm}
\end{figure}

 \noindent Consider Fig. (\ref{setup}), where the channel ${\cal E}$ has a  pure invariant state $|0\ra$. This channel transmits its invariant state undisturbed and distorts or de-coheres any other state. If we project every input  states onto the invariant state, it certainly passes through the channel with perfect fidelity, however on the average only half of the particles are projected onto this invariant state and the rest do not find the chance to enter the channel. Moreover any coherence in the input states is lost in the output state.  On the other hand if we allow all the particle to pass through the channel, then  the average fidelity is lowered. Therefore a compromise should be made between a high input-output average fidelity and a high passage probability which are opposing each other. We do this by a tuned weak measurement  with two elements
 \begin{equation} \label{W0}
 	W_{0}=|0\rangle\langle 0|+\sqrt{p}|1\rangle\langle 1|,
 \end{equation}
 and
 \begin{equation} \label{W1}
 	{W_{1}}=\sqrt{1-p}|1\rangle\langle 1|.
 \end{equation}
 {\bf Remark:} For a channel with invariant state $|\lambda\ra$ the corresponding weak measurement is defined by replacing $|0\ra$ by $|\lambda\ra$ and $|1\ra$ by $|\lambda^\perp\ra$. \\

 \noindent As explained in the previous section, a general channel ${\cal E}^\lambda$ (with invariant state $|\lambda\ra$) is  unitarily related to the channel ${\cal E}$ (with $|0\ra$ as the invariant state). This implies that the pre-weak measurement is also unitarily related for the  channel leading to the same amount of increase of the average fidelity. Henceforth, we restrict ourselves to  the most general channel with invariant state $|0\ra$ described in previous section. \\

  \noindent After the pre-measurement, there are two output states, namely  $\rho_0:=\frac{W_0\rho W_0^\dagger}{Tr(W_0\rho W_0^\dagger)}$ which passes through the channel ${\cal E}$ with  probability $P_0=Tr(W_0\rho W_0^\dagger)$  with little distortion, and   $\rho_1:=\frac{W_1\rho W_1^\dagger}{Tr(W_1\rho W_1^\dagger)}=|1\ra\la 1|$ which is blocked. The compromise should be made between two opposing factors, namely the passage probability which is an increasing function of $p$ and the average fidelity between the input and output states of the channel ${\cal E}$ which is a decreasing function of $p$.  The task is to find the optimum value of this parameter. \\


\subsection{Performance of the protocol}\label{fidelity}
Let a general (not necessarily trace-preserving) channel $\Phi$ has the Kraus operators $K_i$,
\be\label{Phi}
{\Phi(|\psi\rangle\langle \psi|)}=\sum_iK_{i}|\psi\rangle\langle\psi|K_{i}^{\dagger}.
\ee
For this channel, the average input-output fidelity and the average passage  probability (i.e. the probability of having an output state) are respectively given by
\ba \label{Fpi}
F({\Phi})&=&\int d\psi Tr\big(|\psi\rangle\langle\psi|\Phi(|\psi\rangle\langle\psi|)\big)=\sum_i\int d\psi  \  |\la \psi|K_i|\psi\ra|^2\cr
&=&  Tr\big [\sum_i(K_i^\dagger \otimes K_i)\int d\psi |\psi^{\otimes 2}\ra\la \psi^{\otimes 2}|\big],
\ea
and
\ba \label{PSP}
P(\Phi)&=&\int d\psi Tr\big({ \Phi}(|\psi\rangle\langle\psi|)\big)=\sum_i \int d\psi \ \la \psi|K_{i}^{\dagger}K_{i}|\psi\rangle\cr
&=&Tr\big(\sum_i K_i^\dagger K_i\int d\psi|\psi\ra\la \psi|\big),
\ea
in which $d\psi$ is the uniform integral over the Bloch sphere.
Then the normalized  average fidelity is defined by the ratio of these two quantities
\be\label{definiatin of a.f.}
{F_n}({ \Phi}):=\frac{\sum_i \int d\psi \  |\la \psi|K_i|\psi\ra|^2}{\sum_i \int d\psi \ \la \psi|K_{i}^{\dagger}K_{i}|\psi\rangle}.
\ee
 \noindent By this normalization we are effectively calculating the input-output fidelity only for those input states which are permitted to pass through the combined channel, i.e. the weak measurement and the main channel.
We now have a figure of merit which can be calculated exactly.
\noindent The right hand side of equations (\ref{Fpi}, \ref{PSP})  can be recast in a more convenient form if we note that $\la S\ra:=\int d\psi S=\frac{1}{4\pi }\int d(\cos\a) d\beta S $, and use the results
$
\la \cos^2\frac{\a}{2}\ra= \la \sin^2\frac{\a}{2}\ra=\frac{1}{2}, \
\la \cos^4\frac{\a}{2}\ra= \la \sin^4\frac{\a}{2}\ra= 2\la \cos^2\frac{\a}{2}\sin^2\frac{\a}{2}\ra=\frac{1}{3}$.
Here we assumed that any pure state on the Bloch sphere is given by two real parameters $\alpha$ and $\beta$. This leads to
\be
\int d\psi |\psi\ra\la \psi|=\frac{I}{2},\h \int d\psi |\psi^{\otimes 2}\ra\la \psi^{\otimes 2}|=\frac{I+S}{6}.
\ee
where $S$ indicates the swap operator. Therefore, we find in compact form the passage probability and the average fidelity of any qubit channel as follows:
\be \label{Pn}
P(\Phi)=\frac{1}{2}Tr(\sum_i K_i^\dagger K_i),\h F(\Phi)=\frac{1}{6}{Tr\big(\sum_i (K_i^\dagger\otimes  K_i)(I+S)\big)},
\ee
which leads to
\be \label{Fn}
F_n(\Phi)=\frac{1}{3}\frac{{Tr\bigr(\sum_i (K_i^\dagger\otimes  K_i)(I+S)\big)}}{Tr(\sum_i K_i^\dagger K_i)}.
\ee

\noindent
It is now a matter of direct substitution to insert the explicit form of the operators $K_i= A_iW_0$ into (\ref{Pn}, \ref{Fn}) and obtain

\be\label{KKK}
\sum_{i}K_i^\dagger K_i=\left(\begin{array}{cc} 1&0\\ 0&p\end{array}\right),
\ee
and
\be \label{PC}
\sum_{i}K_i^\dagger \otimes K_i=
\left(\begin{array}{cccc} 1&0&0&0\\ 0 & \sqrt{p}y_0&0&0\\ 0&p x^2&\sqrt{p}y_0& p xye^{-i\frac{\phi}{2}}\cos\theta\\ 0& pxye^{i\frac{\phi}{2}}\cos\theta & 0 & p(1-x^2)\end{array}\right),
\ee
where we have used the constraint $y_0^2+x^2+y^2=1$.\\

\noindent This then leads from (\ref{Pn}) to the passage probability of $P_{pass}=P({\cal E}\circ W)=\frac{1}{2}(1+p)$ and from (\ref{Fn}) to the following expression for the figure of merit
\be
F_n^{(p)}:=F_n({\cal E}\circ W)=\frac{1}{3}\frac{2+2p+2y_0\sqrt{p}-px^2}{1+p}.
\ee
When $p=1$, this gives the average fidelity for the original channel ${\cal E}$ which we denote by $F_n^{(1)}$:
\be
F_n^{(1)}:=F_n({\cal E})=\frac{2}{3}+\frac{1}{6}(2y_0-x^2).
\ee

\noindent The improved fidelity finds its maximum value for specific values of  $p$ which is found from scrutiny of the function $F_n^{(p)}$ and its derivatives. It turns out that the optimal value of $p$ is given by 

\be\label{ni0}
p_{opt} = \left\{\begin{array}{lr}
	\frac{x^4 + 2 y_0^2 - \sqrt{x^8 + 4 x^4 y_0^2}}{2 y_0^2}, & \ \ \  y_0>0\\
	0, & \ \ \ y_0\leq 0
\end{array}\right.
\ee

\noindent Inserting this value back into $F_n^{(p)}$ will give the optimum value of average fidelity as a function of $x$ and $y_0$. The difference between this improved fidelity and the original fidelity is shown in figure (\ref{avefid}), where it is seen that the largest amount of improvements are obtained when $y_0$ is negative. 
 It is also instructive to compare these two average fidelities for special subclasses of the channels, namely those which correspond to the equator and the two meridians on figure (\ref{S2fig}). \\

\noindent {\bf Decohering channels (${\cal E}_{Dec}$): } In this case we have $x=0$ and
\be
F_n^{(p)}-F_n^{(1)}=\frac{-y_0(1-\sqrt{p})^2}{3(1+p)}.
\ee
which shows an increase of the average fidelity  only for half of the channels which have a negative parameter $y_0$. The optimal value of the weak-measurement parameter depends on the sign of $y_0$. For $y_0<0$, we have $p^{opt}=0$, corresponding to a projective measurement and for $y_0\geq 0$, we have $p^{opt}=1$ corresponding to no measurement.  \\

\noindent {\bf Collapsing channels (${\cal E}_{C}$):}
In this case we have $y_0=0$ and
\be
F_n^{(p)}-F_n^{(1)}=\frac{x^2(1-p)}{6(1+p)},
\ee
which is always positive, showing that the weak measurement always increases the average fidelity and $p^{opt}=0$ corresponding to a projective measurement. \\

\noindent {\bf Amplitude Damping channels (${\cal E}_{AD}$): }
In this case, we have $y=0$ and $x^2=1-y_0^2$. Taking into this domain, one can find 
$p^{opt}$,  $F_n^{(1)}$ and $F_n^{p}$ as a function only of $x$. The analytical expression for these are not so transparent, but in figure (\ref{avefid}), these channels correspond to the boundary of the colored region, where it is clear how the average fidelity is improved as a function of $x$. \\

\noindent Finally note that from (\ref{Pn}) and (\ref{KKK}), the average probability of passage is $P(\Phi)=\frac{1}{2}(1+p^{opt})$ which is always greater than $\frac{1}{2}$. Therefore by repeating the process approximately twice, one can achieve a high fidelity with this process.

\begin{figure}
	\centering
	\hspace*{0cm}
	\includegraphics*[scale=0.3]{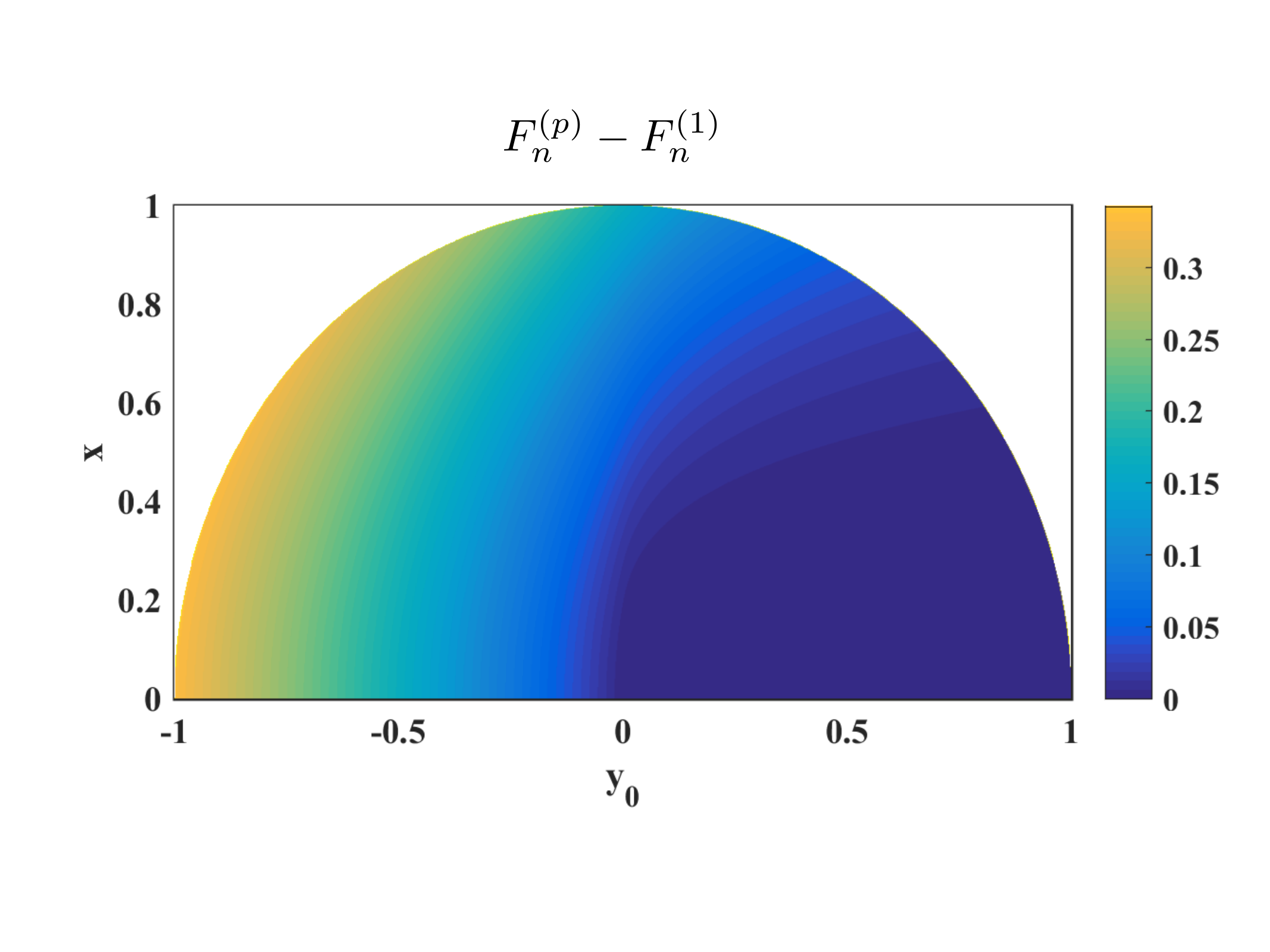}
	\vspace*{-1.5cm}
	\caption{The variations of $F_n^{(p)}-F_n^{(1)}$ as  functions of $x$ and $y_0$ in the allowable region for these parameters $x^2+y_0^2\leq 1.$   It is seen that generally for negative values of $y_0$, the increase of average fidelity is more pronounced. The value of $p$ is the optimal value given in Eq. (\ref{ni0})}\label{avefid}
		\end{figure}

\begin{figure}
	\centering
	\hspace*{0cm}
	\includegraphics*[scale=0.3]{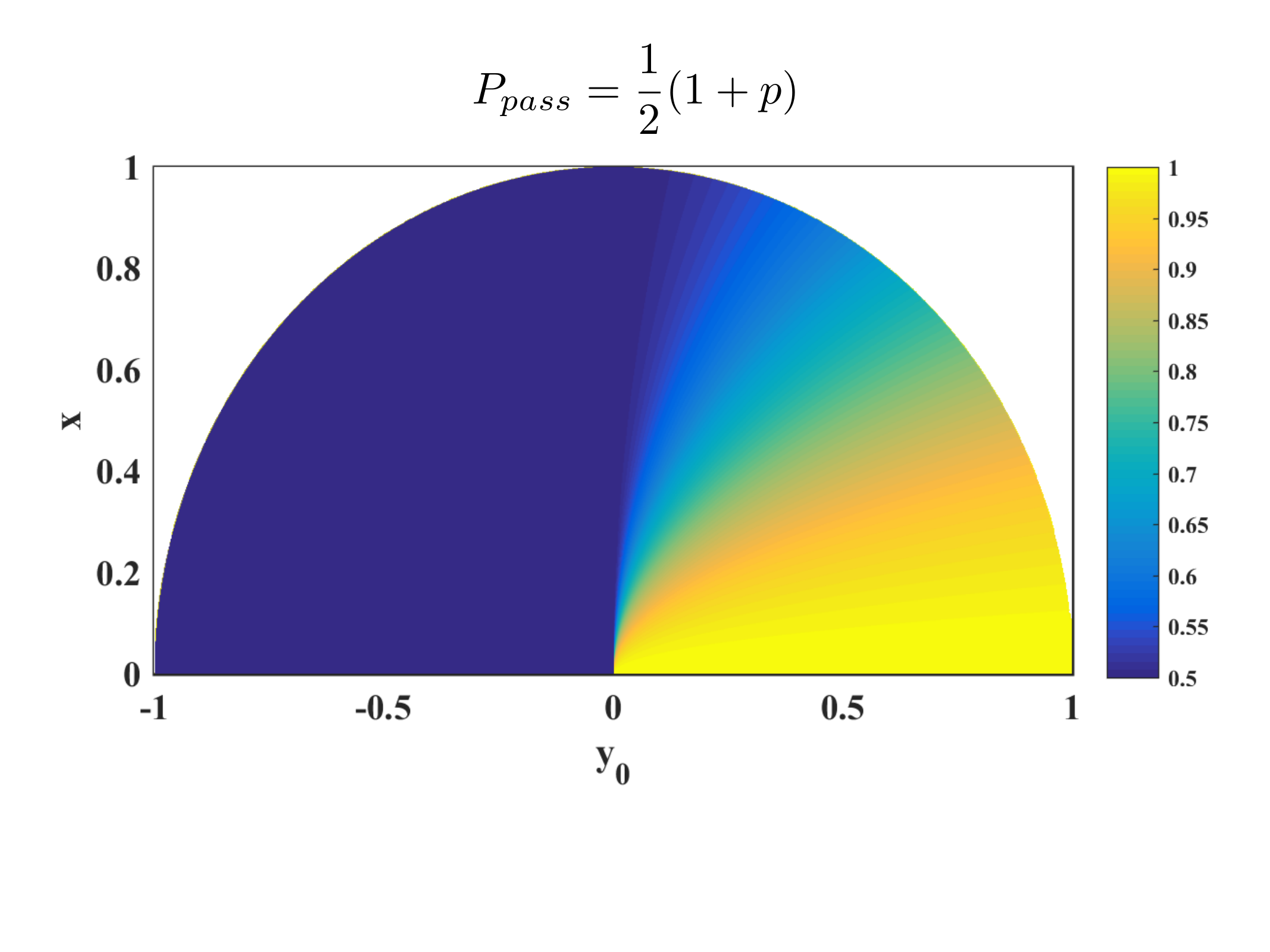}
	\vspace*{-1.5cm}
	\caption{ The passage probability as a function of $x$ for various values of $y_0$.  The value of $p$ is the optimal value given in Eq. (\ref{ni0})}\label{avefid}
\end{figure}

\section{Conclusion}\label{conclusion}
What we have done in the present work is in line with a recent trend of works \cite{revers3, , state1, entangle1, discord3} which seem to be   practical procedures for combating de-coherence.  These works have been mainly concerned with the Amplitude Damping channel and for protecting specific kinds of states, i.e. entangled states or states which have discord. Here we have generalized these works to a large class of channels and have also considered general input states and average fidelity between input and output. We have shown that by suitably tuning the parameter of a weak measurement, the average fidelity can be increased. 
 To this end, we have focused on one important aspect of quantum channels which have an invariant pure state.  \\

 \noindent As a byproduct, we have classified all qubit channels with one invariant pure state and have shown that all such channels are unitarily equivalent to a channel represented by at most three Kraus operators embodying four real parameters.\\
 
 \noindent 
   Several questions remain to be investigated in the future. The most notable one will be to classify all higher dimensional channels with one or more invariant pure states. For arbitrary dimensions, in view of the quadratic increase of the number of parameters and  lack of  geometric pictures like the Bloch sphere, this is a highly nontrivial problem, but it can possibly be done for low dimensions. The idea of having one or more invariant pure states, can itself lead to a partial classification of higher dimensional channels which can in turn be pursued further in the context of the present paper, namely state protection from de-coherence.  One can also see how this scheme can protect quantum correlations of two or more qubit states, if applied to one of the shares. Finally it is an interesting question to see what happens if we consider also mixed states as input states to the channel. This is not a conceptually  difficult problem and will presumably will not change the protocol itself. However it is not so easy to find analytical expressions for average fidelity since in this case the fidelity is a non-linear function of the input state  $\rho$. Instead one  can invoke the convexity property of fidelity according  to which for any channel $\Phi$ and any decomposition of $\rho=\sum_i p_i|\psi_i\ra\la \psi_i|$, one has
   \be
   F(\rho, { \Phi}(\rho))\geq \sum_i p_i F(|\psi_i\ra\la \psi_i|, \Phi(|\psi_i\ra\la \psi_i|)).
   \ee
   This then means that instead of taking the average fidelity as a figure of merit, one should take the minimum fidelity which according to the above inequality is always satisfied for a pure state. However this will take us too far from our study in this paper and we prefer to deal with this later on.

\section{Acknowledgement}  We thank the anonymous referees for their valuable comments which helped to significantly improve the presentation of our results.  This research was partially supported by a grant no. G98024071 from Iran National Science Foundation. This research was done in O. Heibati's sabbatical leave in Sharif University of Technology as part of the requirement for completion of her PhD thesis.


 \vspace{1cm}\setcounter{section}{0}
 \setcounter{equation}{0}
 \renewcommand{\theequation}{C-\roman{equation}}


\end{document}